\documentclass[12pt]{iopart}
\usepackage{graphicx}
\usepackage{amssymb}

\newcommand{\bra}[1]{\langle #1 | \,}
\newcommand{\ket}[1]{\, | #1 \rangle}
\newcommand{\braket}[2]{\langle #1 | #2 \rangle}

\newcommand{\iim}{\mathrm{i}}
\newcommand{\be}{\begin{equation}}
\newcommand{\ee}{\end{equation}}
\newcommand{\bea}{\begin{eqnarray}}
\newcommand{\eea}{\end{eqnarray}}
\newcommand{\besa}{\begin{subeqnarray}}
\newcommand{\eesa}{\end{subeqnarray}}
\newcommand{\bean}{\begin{eqnarray*}}
\newcommand{\eean}{\end{eqnarray*}}

\newcommand{\had}{\hat{a}^{\dagger}}
\newcommand{\ha}{\hat{a}}

\begin{document}

\title{Lattice oscillator model, scattering theory and a many-body problem}

\author{Manuel Valiente}
\address{Lundbeck Foundation Theoretical Center for Quantum System Research, Department of Physics and Astronomy, Aarhus University, DK-8000 Aarhus C, Denmark}

\date{\today}

\begin{abstract}
We propose a model for the quantum harmonic oscillator on a discrete lattice which can be written in supersymmetric form, in contrast with the more direct discretization of the harmonic oscillator. Its ground state is easily found to be annihilated by the annihilation operator defined here, and its excitation spectrum is obtained numerically. We then define an operator whose continuum limit corresponds to an angular momentum, in terms of the creation-annihilation operators of our model. Coherent states with the correct continuum limit are also constructed. The versatility of the model is then used to calculate, in a simple way, the generalized position-dependent scattering length for a particle colliding with a single static impurity in a periodic potential and the exact ground state of an interacting many-body problem in a one-dimensional ring.

\end{abstract}


\maketitle

\section{Introduction}
The quantum harmonic oscillator is a paradigmatic model with applications in all
branches of physics, too numerous to be counted. Due to the special structure
of its Hamiltonian, it is possible to obtain its eigenstates exactly in several
different ways. Perhaps the most celebrated one is the algebraic solution by means
of the creation-annihilation (ladder) operators \cite{reviewSUSY}, which is by
far the simplest and most elegant and, moreover, it is the first step towards
field quantization \cite{Messiah}. 

For more complicated problems the use of numerical methods
becomes necessary. One of the most popular techniques is the finite-difference
discretization, which is often employed in high-energy physics to obtain
non-perturbative results \cite{Rothe}. However, this method has severe
problems, since the symmetries of the original problem are usually lost on the
lattice and can only be recovered once the continuum limit is taken which, in
practice, is numerically impossible. As an
important example, the lattice harmonic oscillator, though it
can be written as the Mathieu equation in quasi-momentum space \cite{Gallinar,RevMattis},
is not factorizable and not even its ground state can be obtained in closed
form. This fact is very disappointing, since then the supersymmetric (SUSY) structure of the system
is not transparent -- in fact, not even present -- until the continuum limit is taken.

In this article, we construct a model for the lattice harmonic oscillator
which has a correct continuum limit. Its Hamiltonian is shape invariant
\cite{reviewSUSY} and,
though the excitations cannot be accessed analytically, its ground state is
exactly solvable for any value of the oscillator frequency and the lattice
spacing. The excitations can, however, be obtained by solving an equation
which is analogous to the
Hermite equation. We propose then a definition of coherent states, finding
that their correct continuum limit cannot be obtained if they are defined as
eigenstates of the lattice annihilation operators, so their definition has to
be given in terms of the displacement operator. Our model is completely analogous
to that for a single particle in a periodic potential, and we use it to
calculate the lowest band zero energy scattering length in a particle-impurity
collision. We then make further use of the analogy of
the model with a many-body system with anharmonic interactions on a finite
ring which can be solved exactly for the ground state.


\section{Position and momentum operators on the lattice}

We define the following operators in quasi-momentum space as the momentum
($\hat{p}$) and position ($\hat{x}$) operators,
\begin{eqnarray}
\hat{p} &\equiv \frac{\hbar}{d}\sin{kd}\label{p}\\
\hat{x} &\equiv \iim\frac{\partial}{\partial k},\label{x}
\end{eqnarray}  
where $d$ ($>0$) is the lattice spacing and $k\in(-\pi/d,\pi/d]$ is the quasi-momentum. The
operators $\hat{p}$ and $\hat{x}$ are constructed in analogy
with their continuous space counterparts. Note that $\hat{x}$ coincides with its continuum analog while $\hat{p}= \hbar k + O(d^2)$ as $d\to 0$, and therefore their continuum limits are correctly described. We can write the lattice
analog of the harmonic 
oscillator annihilation $\hat{a}$ and
creation $\hat{a}^{\dagger}$ operators as
\begin{eqnarray}
\hat{a} &=  (\hat{X}+\iim\hat{P})/\sqrt{2}\\
\hat{a}^{\dagger} &=  (\hat{X}-\iim\hat{P})/\sqrt{2},
\end{eqnarray} 
where the quadrature operators are defined as 
\begin{eqnarray}
\hat{X}&\equiv
(m\omega/\hbar)^{1/2}\hat{x},\\
\hat{P}&\equiv (m\hbar\omega)^{-1/2}\hat{p}.
\end{eqnarray}
However, by using the lattice operators of
Eqs. (\ref{p}) and (\ref{x}) we see that 
$[\hat{X},\hat{P}]= \iim \cos(kd)$
 and $[\hat{a},\hat{a}^{\dagger}]=\cos(kd)$. In other words, the canonic commutation relations are valid up to a
factor of $\cos(kd)$. In the limit of small lattice spacing, we obtain the
correct commutation relation of the continuous space case $\lim_{d\to 0} [\hat{X},\hat{P}] =
\iim$. The commutation relation $[\hat{X},\hat{P}]$  yields a generalized
uncertainty principle (GUP) \cite{Elias} of the form
\be
\Delta X\Delta P \ge \frac{1}{2} |\langle \cos(kd) \rangle|,\label{GUP}
\ee
which resembles the GUP of systems with a minimal length
\cite{Hossenfelder}. The GUP (\ref{GUP}) does not imply any minimal
$\Delta X$ (it can be zero). However, we have a maximal dispersion for the
momentum,
\be
\Delta P \le \Delta P_{\mathrm{max}}= \sqrt{\frac{\hbar \pi}{m\omega d^3}},
\ee 
which is infinite in the continuum limit, as it should.

\section{The lattice harmonic oscillator}

If we wish to construct a lattice theory for the harmonic oscillator having
a similar structure as its continuum limit, we have to consider the operator
\be
\hat{N}\equiv \had\ha = \frac{1}{2} (\hat{P}^2 + \hat{X}^2 + \iim
[\hat{X},\hat{P}]).\label{number} 
\ee  
So far, the ``number'' operator, Eq.\ (\ref{number}), has exactly the same
appearance as in continuous space. Its explicit form is given by
\be
\hat{N} = \frac{1}{2}\left(-\frac{m\omega}{\hbar}\frac{\partial^2}{\partial
    k^2} + 
  \frac{\hbar}{m\omega d^2}\sin^2(kd) - 
  \cos(kd)\right).\label{numberSUSY}
\ee
The number operator written in this way looks rather unusual. If we rewrite
$\sin^2(kd) = (1-\cos(2kd))/2$, and perform  Fourier transform to direct lattice space, then we see that the number operator $\hat{N}$
acts as
\begin{eqnarray}
(\hat{N}\psi)(x)&=\frac{1}{2}\big[\frac{m\omega}{\hbar}x^2 \psi(x) 
-\frac{\psi(x+d)+\psi(x-d)}{2}\nonumber \\
&+\frac{\hbar}{2m\omega d^2}\big(\psi(x)-\frac{\psi(x+2d)+\psi(x-2d)}{2}\big)
\big]
\end{eqnarray}
where $n=x/d\in \mathbb{Z}$ are the lattice points. Thus, the number operator 
corresponds to a lattice with nearest-neighbor and next-nearest-neighbor
hoppings with an external harmonic trap, plus a trivial constant. After taking
the continuum limit $d\to 0$, one easily verifies that $\hbar\omega\hat{N}\to
\hat{p}^2/2m + m\omega^2x^2/2 -\hbar\omega/2$. 

\subsection{Ground state and lattice Hermite equation}

From now on we consider the Hamiltonian
\be 
H\equiv \hbar \omega \big(\hat{N}+\frac{1}{2}\big).\label{latticeHam}
\ee
If 
operator $\hat{a}$ annihilates a wave function in $k$-space which is
$2\pi$-periodic,\cite{footnote2} 
then it is the ground state of $H$ with energy 
$E_0= \hbar\omega/2$. Equation $\hat{a}\psi_0(k) = 0$ is readily solved and the
ground state wave function has the form
\be
\psi_0(k) = \mathcal{N} e^{-\gamma_d} e^{\gamma_d\cos(kd)},\label{ground}
\ee
where $\gamma_d\equiv \hbar/(m\omega d^2)$. 
In the continuum limit, Eq.\ (\ref{ground}) reduces to the
well-known harmonic oscillator ground state,
$\psi_0(k) \sim \exp(-\hbar k^2/m\omega)$. We use this result to verify the
uncertainty principle on the lattice, and find that in the ground state of the
lattice harmonic oscillator, the uncertainty relation is
also minimal,
\be
\Delta X \Delta P = \frac{1}{2} \left|\frac{\bra{\psi_0}\cos(kd)\ket{\psi_0}}{\braket{\psi_0}{\psi_0}}\right|,
\ee
since in general $\Delta X \Delta P \ge
|\langle\big[\hat{X},\hat{P}\big]\rangle|/2$. 
 
Further analogy with the harmonic oscillator in continuous space can be
observed by solving the eigenvalue problem for the number operator
$\hat{N}\psi(k) = \tilde{N}\psi(k)$ with the 
ansatz $\psi(k) = \psi_0(k)\phi(k)$. The eigenvalue problem is then
transformed to the equation
\be
\phi''(k) - 2 \frac{\hbar}{m\omega}\frac{\sin(kd)}{d} \phi ' (k) + 2
\frac{\hbar}{m\omega} \tilde{N} \phi(k)=0\label{Hermite}
\ee
that determines the unknown $\phi(k)$ for which periodic boundary conditions (PBC) $\phi(k+2\pi/d)=\phi(k)$ are assumed.
Note the analogy of Eq.\ (\ref{Hermite}) with the Hermite equation: the naive
substitution $\sin(kd)/d\sim k$, valid for $kd\ll 1$, yields the well-known
continuum limit, in which the eigenvalues $\tilde{N}$ become natural numbers.

\begin{figure}[ht!]
\begin{minipage}[h]{\linewidth}
\centering {\includegraphics[width=\linewidth]{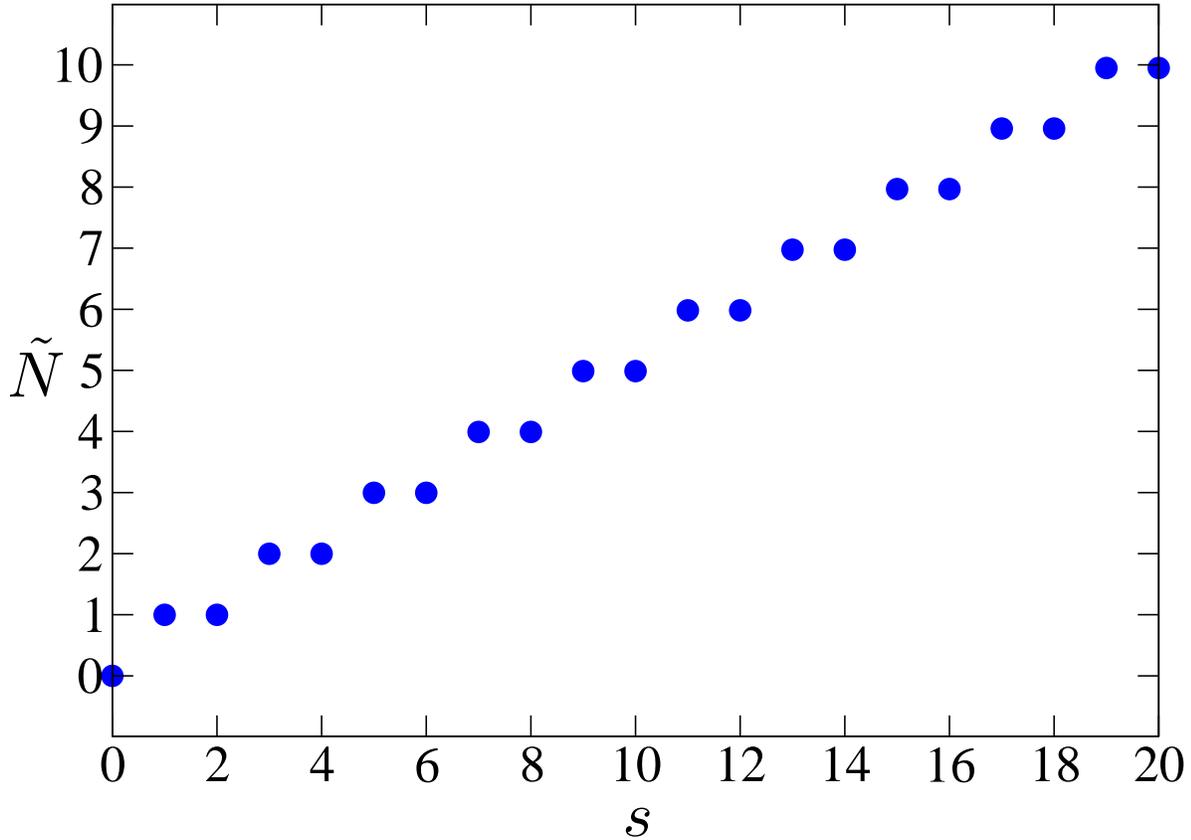}}
\caption[Spectrum of the lattice number operator]{Eigenvalues of $\hat{N}$,
  Eq.\ (\ref{numberSUSY}), for
  $d=\frac{1}{10\sqrt{5}}\sqrt{\frac{\hbar}{m\omega}}$.}
\label{fig:SUSYspectrum}
\end{minipage}
\end{figure}

In Fig. \ref{fig:SUSYspectrum} we plot the low-energy eigenvalues $\tilde{N}$ of
$\hat{N}$, Eq.\ (\ref{numberSUSY}), for a small value of the lattice spacing
$d$. We see that the lowest eigenvalue is indeed zero,
while the rest of the eigenvalues appear to be quasi-degenerate but almost
linearly spaced as $\tilde{N}_{s+2}-\tilde{N}_{s} = 1$. The reason is that, in
direct lattice space, the number operator includes tunneling to nearest and
second nearest neighbors, therefore inducing the quasi-degeneracy, except for
the ground state. The relevant eigenstates for the continuum limit are those
labeled by even quantum numbers $s$ and, in direct lattice
space, appear to be essentially a superposition of the discretized Hermite functions
$\psi_{s/2}(x = 2nd) - \psi_{s/2}(x = - (2n+1)d)$.

\subsection{Coherent states}

It is natural to define now the coherent states for the lattice harmonic
oscillator. First, we try the eigenstates of the
annihilation operator, $\hat{a}\psi_{\alpha} = \alpha \psi_{\alpha}$, with
$\alpha\in\mathbb{C}$. The solution of this equation is 
\begin{equation}
\psi_{\alpha}(k) = \exp\left[-\iim 2\sqrt{\frac{\hbar}{m\omega}}\alpha
  k\right] \psi_0(k) 
\end{equation}
with $\psi_0(k)$ being the ground state, Eq.\ (\ref{ground}). If we assumed $\alpha$ to be \textit{any} complex
number, $\psi_{\alpha}$ would not fulfill the PBC. Hence, if we insist on
$\psi_{\alpha}$  to be
correct, we have no choice but to restrict the values of $\alpha$ to 
$\alpha_j = \frac{1}{2}\sqrt{\frac{m\omega}{\hbar}} j, \hspace{0.2cm}
j\in\mathbb{Z}$, which is a very unsatisfactory answer since the
coherent states would then be restricted to equally spaced real numbers. Therefore the
coherent states $\psi_{\alpha}$ do not present a very
convenient definition. This apparently difficult situation can be resolved in a
rather elegant way, however, relaxing the requirement that the coherent states be
eigenstates of the lattice annihilation operator $\hat{a}$. To this end, we
\textit{define} the coherent states $\Psi_{\alpha}$ as solutions of the
equation $\hat{a}\Psi_{\alpha}(k) = \alpha \cos(kd) \Psi_{\alpha}(k)$, 
\be
\Psi_{\alpha} (k) = \exp\left[-\iim 2 \alpha \sqrt{\frac{\hbar}{m\omega}} \frac{\sin(kd)}{d}\right]\psi_0(k),\label{coherent}
\ee 
which are obviously $2\pi$-periodic for all $\alpha\in \mathbb{C}$, and have
the correct continuum limit.
We can further justify Eq.\ (\ref{coherent}) as a definition since even in the
continuum the coherent states are solutions of
$(\hat{a}+\alpha\iim[\hat{X},\hat{P}])\Psi_{\alpha} = 0$. The only issue we cannot
generalize to the lattice case is the usual form of the displacement operator, since on the
lattice the Baker-Hausdorff formula is not valid due to the commutator
$\big[\hat{a},[\hat{a},\hat{a}^{\dagger}]\big]\ne 0$. Therefore we define here the
displacement (or translation) operator for lattice and
continuum as $\hat{D}(\alpha) = e^{-\iim 2\alpha \hat{P}}$,
which generates \textit{unnormalized} coherent states.

\subsection{Angular momentum}

A major inconvenience of lattice discretizations, if these are introduced
artificially and not due to a true underlying crystal structure, is the
absence of continuous rotational symmetry. To be more concrete, we lack conservation of angular momentum
or, more dramatically, we do not even have a definition of angular momentum on
the lattice!

We propose here a rather simple lattice analog of the angular
momentum. The requirements this operator has to
satisfy are rather relaxed: (i) it should have a correct continuum
limit, (ii) there should be a ground state of some
relevant enough Hamiltonian with a well defined ``angular momentum'' on the
lattice and, (iii) the lattice angular momentum cannot commute with the
lattice Hamiltonian.   
Requirement (iii) is easy to satisfy: take \textit{any} Hamiltonian which respects the
symmetries of the lattice. We discuss now how
(i) and (ii) can be met. Consider first a two-dimensional (2D)
oscillator with Hamiltonian 
$H^{(\mathrm{2D})} = \hbar\omega\big[\hat{N}_1+\hat{N}_2 +1]$, 
with $\hat{N}_i = \had_i\ha_i$. We define the lattice angular momentum in 2D, in analogy
with the continuum, as
\be
\hat{L}\equiv \hat{x}_1\hat{p}_2-\hat{x}_2\hat{p}_1 = \iim
(\ha_1\had_2-\had_1\ha_2),\label{angular}
\ee   
and we see that condition (i) is fulfilled. 
As promised, requirement (ii) is automatically satisfied by
the ground state of $H^{(\mathrm{2D})}$, $\psi_0^{(\mathrm{2D})}(k_1,k_2)=\psi_0(k_1)\psi_0(k_2)$,
with $\psi_0(k_i)$ defined in Eq.\ (\ref{ground}), and by the ground state of
the 2D free particle Hamiltonian $H_F = -2J\sum_{k_1,k_2}(\cos(k_1 d)+\cos(k_2
d) )\ket{k_1,k_2}\bra{k_1,k_2}$,
$\psi(k_1,k_2) = \delta(k_1)\delta(k_2)$, both having 
angular momentum $L=0$ as a good quantum number. It must be noted that lattice angular momentum operators
have been defined in the context of rotating gases in an optical lattice
in \cite{Holland}, but with such definition requirement (ii) is no longer
satisfied for the ground state of the lattice oscillator.
We now show by explicit calculation that the
angular momentum on the lattice is in general not a conserved quantity
\begin{eqnarray}
[\hat{L},\hat{N}_1+\hat{N}_2] = &\iim
\big[\cos(k_1d)\ha_1\had_2+\ha_2\had_1\cos(k_1d)\nonumber \\
&-\ha_1\had_2\cos(k_2d) -
\cos(k_2d)\ha_2\had_1\big],
\end{eqnarray}
which, as expected, is non-zero, but vanishes in the continuum limit. It
must be noted now that the lattice angular momentum operator,
Eq.\ (\ref{angular}), can be used as a definition not only for the model
discussed here, but for \textit{any} tight-binding
lattice model even without next-nearest-neighbor hopping.

\section{Application to impurity scattering in a periodic potential}
\begin{figure}[ht!]
\begin{minipage}[h]{\linewidth}
\centering {\includegraphics[width=\linewidth]{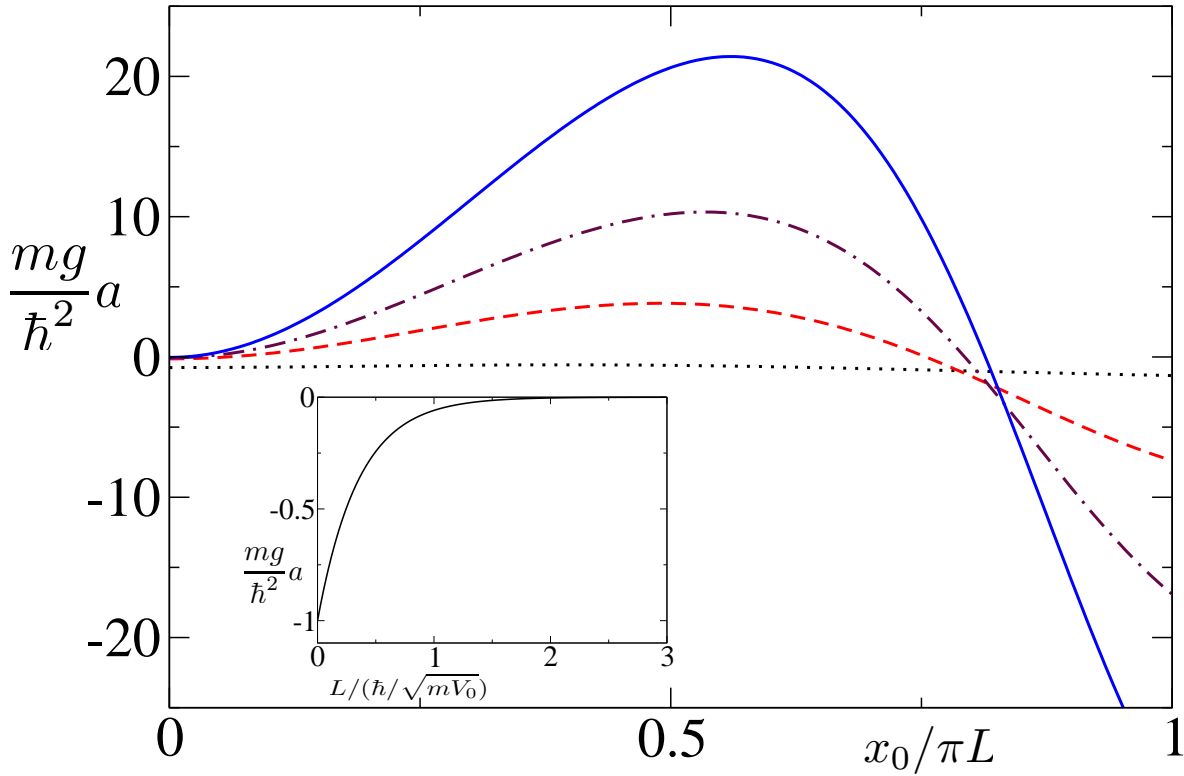}}
\caption[Generalized scattering length in periodic potentials]{Scattering lengths for $mV_0L^2/\hbar^2=10^{-3}$, $1/2$, $1$ and
  $3/2$ (at $x_0=0$ 
  from bottom to top). Inset: $a(0)$ as a function of $L$.}
\label{fig:scatlength}
\end{minipage}
\end{figure}


The model presented here can be applied to construct completely different systems
and obtain some of their properties exactly. As a first application, let us
consider a single particle moving on the real line. It is readily
verified that the Hamiltonian with the periodic potential,
\be
V(x) = V_0 \sin^2(x/L)-\frac{\hbar}{L}\sqrt{\frac{V_0}{2m}} \cos(x/L)
\ee
has a ground state $\psi_0(x) = \exp[\lambda_L \cos(x/L)]$, with
$\lambda_L=\sqrt{2mV_0}L/\hbar$, since the potential and kinetic energy
operators are dual to those for the lattice Harmonic oscillator in
quasi-momentum space. We consider
a single static impurity located at $x_0\in (-\pi L,\pi L]$ (this is the central
site, and by translation applies to an impurity at any site), with zero range
interaction potential $V_g(x) = g\delta(x-x_0)$, and we show how to get the
low-energy scattering properties of the system in a very simple manner. First
we notice that, since $V$ has a purely continuous spectrum and the impurity is
immobile, upon collision the incident waves can only acquire a phase shift. Therefore, for low-energy scattering we only need the periodic
($\psi_0$)  and aperiodic (which we call $\psi_I$) solutions for zero
energy. The aperiodic solution centered at the first site is given by
\be
\psi_I(x) = \psi_0(x)\int^x dx e^{-2\lambda_L \cos(x/L)}\equiv \psi_0(x)\Phi(x).
\ee    
This aperiodic solution is clearly antisymmetric and it holds that $\Phi(x) =
\beta x +\phi_p(x)$, where $\phi_p(x+2\pi L)=\phi_p(x)$. Recall that \textit{without} the periodic potential, this solution corresponds to setting
$\phi_p\equiv 0$, and the scattering length $a_0$ of a static Dirac delta impurity
is defined \cite{LSSY} 
by the zero-energy solution $f_0(x) = 1-|x|/a_0$. Clearly, $f_0$ is the sum of
the periodic (free) solution and the aperiodic (unnormalizable) solution, with
 the inverse scattering length as a coefficient. In analogy to the free space
situation, we define a position-dependent scattering length $a(x_0)$, $x_0\in
(-\pi L,\pi L]$, in terms of the zero-energy solution
\be
f(x;x_0) = \psi_0(x) -\frac{|\psi_I(x)-\psi_I(x_0)|}{a(x_0)},
\ee
which, written in this way, satisfies the boundary condition
\be
f'(x_0^+)-f'(x_0^-) = 2mgf(x_0)/\hbar^2
\ee
 imposed by the Dirac
delta, if 
\be
a(x_0) = -\frac{\hbar^2}{mg}
\left[\frac{\left(\psi_0\Phi\right)'}{\psi_0}\right]_{x=x_0}
\ee
for $x_0\in (-\pi L,\pi L]$, and $a(x_0+n2\pi L)=a(x_0)$, with
$n\in\mathbb{Z}$. In the simplest case of $x_0=0$, the scattering length is
shown to have the form $a(0)=-\hbar^2 e^{-2\lambda_L}/(mg)$. If $x_0\ne 0$ it
has to be calculated numerically. In Fig. \ref{fig:scatlength} we plot the
scattering length $a(x_0)$, showing how strongly it depends on the position
of the impurity. The scattering length never diverges (there is no resonance),
but $mg a(x_0)/\hbar^2$ can actually become positive and indeed very large with
increasing 
$V_0$ at $x_0\ne 0$, even though the corresponding free-space scattering
length (we assume $g>0$) is
negative. This means that interactions in a periodic potential can effectively
change both
quantitative and qualitatively, depending on where the scattering takes place.  

\section{A many-body system}
\begin{figure}[ht!]
\begin{minipage}[h]{\linewidth}
\centering {\includegraphics[width=\linewidth]{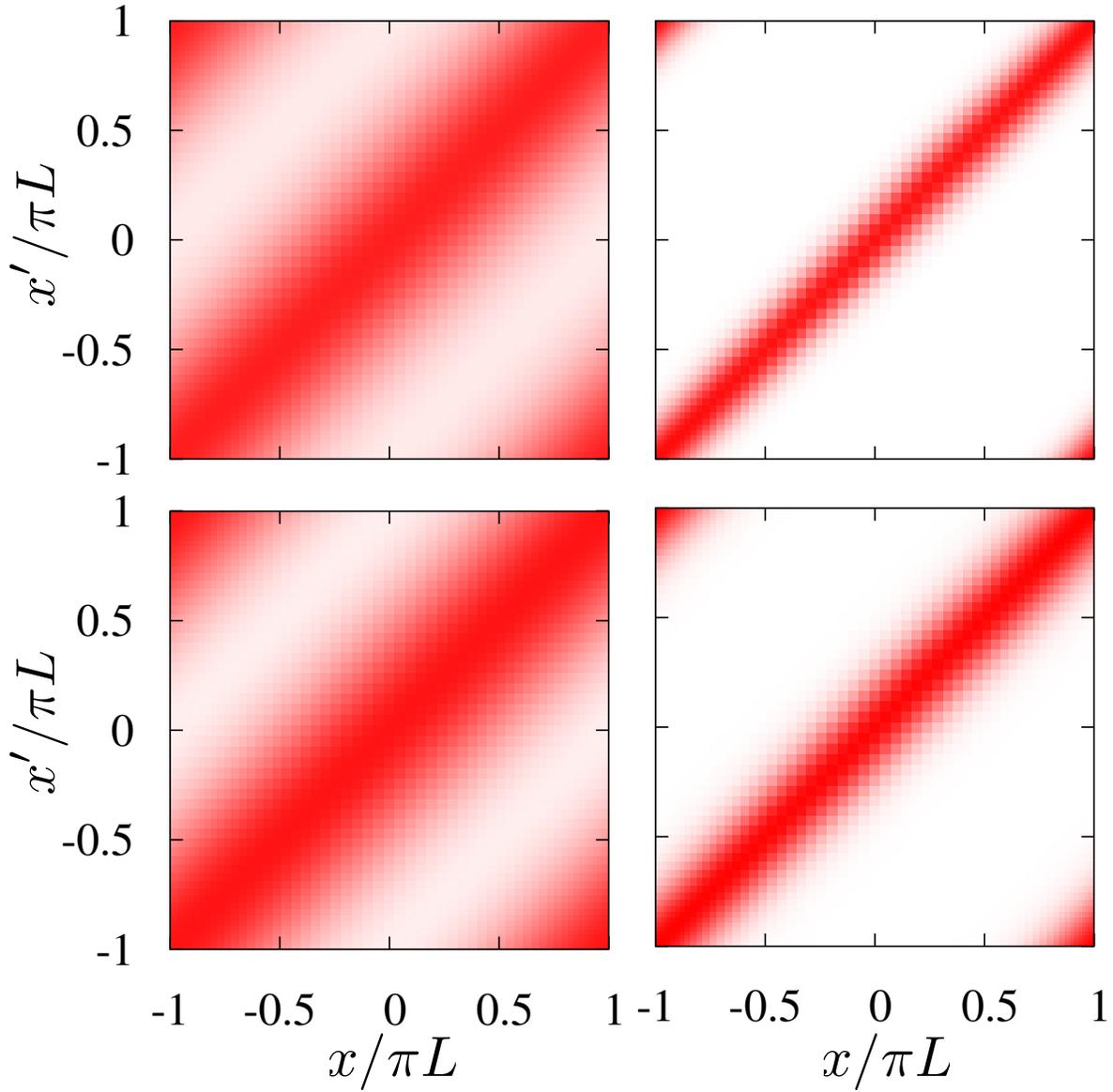}}
\caption[Pair correlation functions on a ring]{Pair correlation functions
  $\rho(x,x')$; the clearer the color the lower its value. Top $N=4$ particles, bottom $N=2$ particles. Left $\lambda
L^2 = 0.1$, right $\lambda L^2 = 1$. All values are normalized to the peak.}
\label{fig:densities}
\end{minipage}
\end{figure}
As a second application, we
construct a many-body Hamiltonian
of interacting particles on a finite ring 
whose ground state can be obtained in closed form. We consider $N$ particles on
a ring of length $2\pi L$. The position of 
particle $i\in \{1,\ldots,N\}$ is denoted by $x_i\in (-\pi/L,\pi L]$ and its
momentum by $p_i = -\iim\hbar \partial/\partial x_i$, and for all the functions
involved we use PBC. We consider the following
Hamiltonian 
\begin{eqnarray}
H&=\frac{\hbar^2}{2m} \sum_{i=1}^N \hat{A}_i^{\dagger}\hat{A}_i,\label{manybodyHam}\\ 
\hat{A}_i &= \frac{\partial}{\partial x_i} + \lambda L\sum_{j=1,  j\ne i}^N  \sin(x_{i,j}/L),
\end{eqnarray}  
where $x_{i,j}=x_i-x_j$.
With these definitions, the many-body Hamiltonian (\ref{manybodyHam}) is
$2\pi$-periodic. In the limit of an infinitely long
ring, $L\to \infty$, we are left
with $N$ particles interacting via pairwise harmonic potentials. However, for
any finite-size ring the interactions are anharmonic and contain three-body
terms. 

Since the Hamiltonian $H$, Eq.\ (\ref{manybodyHam}), is the sum of semi-positive
operators, it follows that $H\ge 0$. Hence, if there exists a non-singular
periodic  function $\psi_0$ which is annihilated by all $A_i$,
$i=1,\ldots,N$, then it is the ground state of $H$ and its eigenenergy is
zero. The set of $N$ equations $\hat{A}_i\psi_0 = 0$ is easily shown to be
satisfied by the wave function
\be
\psi_0(x_1,\ldots,x_N)= \mathcal{N}\prod_{i<j = 1}^N \exp\Big[2\lambda L^2
\cos(x_{i,j}/L)\Big],\label{manybodyground}
\ee
with $\mathcal{N}$ the normalization constant. It is remarkable that for any
$L< \infty$, the ground state (\ref{manybodyground}) is square integrable, even
if $\lambda <0$, but in taking the limit of $L\to \infty$ this will no longer be
true. 

In Fig. \ref{fig:densities} we plot some ground state pair correlation
functions, defined as 
\be
\rho(x,x')\propto\int_{\Omega} dx_3\ldots
  dx_N 
  |\psi_0(x,x',\ldots,x_N)|^2,
\ee
where $\Omega\equiv (-\pi/L,\pi/L]^{N-2}$. We note that as $\lambda L^2$($>0$)
becomes larger, the particles tend to be tighter co-localized, which also
happens for increasing number of particles.    
 
\section{Conclusions}

We have constructed a lattice model of the harmonic oscillator with a correct
continuum limit whose properties, especially in the ground state, are the
perfect analogous of those in continuous space. We have also defined lattice
coherent states and a lattice ``angular momentum'' in terms of the creation
and annihilation operators of the model. By establishing connections
with other systems, we were able to describe low-energy
scattering in a periodic potential and an anharmonically interacting many-body
system. These results are relevant for lattice simulations, cold
collisions, many-body theory and quantum information \cite{marcelo1,marcelo2}.

\section*{Acknowledgements}
I thank Daniel C. Mattis for interesting discussions and suggestions, and David Petrosyan for useful comments on a previous version of the manuscript. I am grateful to Klaus M{\o}lmer for encouragement and support. The author acknowledges financial support from a Villum Kann Rasmussen block scholarship.


\begin{thebibliography}{99}
\bibitem{footnote2}
For simplicity, we will call a function
$2\pi$-periodic when it is actually $2\pi /d$-periodic.
\end{thebibliography}
\end{document}